\documentclass[debug,overfull]{epl}

\title{Dynamic regimes of hydrodynamically coupled \newline self-propelling particles}
\author{Isaac Llopis\inst{1} \and Ignacio Pagonabarraga\inst{1}}

\institute{                    
  \inst{1} Departament de F\'{\i}sica Fonamental, Universitat de
Barcelona, C. Mart\'{\i} i Franqu\'es 1, 08028 Barcelona, Spain\\
}


\begin{document}

\maketitle

\begin{abstract}
We analyze the collective dynamics of  self-propelling  particles (spps) which move at small Reynolds numbers  including the hydrodynamic coupling to the suspending solvent through numerical simulations. The velocity distribution functions  show marked deviations from Gaussian behavior at short  times, and the mean square displacement at long times shows a transition from diffusive to ballistic motion for appropriate driving mechanism at low concentrations. We discuss the structures the  spps form at long times and how they correlate to their dynamic behavior.
\end{abstract}

\section{Introduction}
A great variety of  microorganisms, such as bacteria, have the ability to displace in a fluid medium. Organisms have developed a
number of swimming strategies using  for example  flagella, cilia, or
surface waves~\cite{Bray}. These individuals always produce the force
needed to generate their motion. As a result, they are out of thermal
equilibrium, which makes their motion to differ  qualitatively from
the motion of suspensions of passive particles, such as
colloids~\cite{colloids}. Active motion is not restricted to living
organisms, and is shared by a number of biomolecules; e.g.  membrane
proteins  which move or rotate as a result of the energy they get from
ionic pumps.  In all these situations, propulsion  induces flow in the
embedding solvent leading to effective interactions among the active
particles. Although it is known how these interactions affect the
dynamics of passive suspensions, for example colloids, the role they
play in self-propelling  particles (spps) suspensions  remains to be understood.

Despite  substantial effort and progress have improved the
understanding and characterization of self propulsion and how it
emerges from the elementary molecular  mechanisms~\cite{Bray}, the
implications and characterization of this motion on collective scales
remain far less well understood. Although the hydrodynamics of
microorganisms has been addressed~\cite{pedley}, with special emphasis
on bioconvection~\cite{goldstein}, the relation of this hydrodynamic
behavior with the detailed properties of spps remains a challenge. Recent experimental evidence shows the need of a detailed understanding to explain the   collective particle motion at small scales induced by hydrodynamics~\cite{howard},  or to describe beads' diffusion in bacteria suspensions~\cite{libchaber00}. Such insight is also useful to clarify  the  relevance of  spps' orientation on their rheological response~\cite{ramaswamy04}. Theoretical studies have analyzed the role  spps' interactions have in the emergence of collective behavior (such as flocking) and have provided a framework to understand their properties~\cite{toner}, although disregarding hydrodynamic interactions.

In this letter we address the different dynamical regimes which characterize spps suspensions. We will  focus on the relevant
hydrodynamic couplings at the relevant time scales; an analysis analogous to the one carried out to characterize the dynamics of  passive suspensions~\cite{colloids_1}. We will disregard any coupling other than hydrodynamic  and excluded volume to  focus on their specific relevance to collective propulsion. 

\section{Model}
We   will consider a lattice Boltzmann  (LB) model to describe the fluid and address the
appropriate time scales which determine the coupling of the active
particles and the fluid. The state of the fluid  is specified  by the
average number of fluid molecules, $n({\bf  c},{\bf r},t)$, with
velocity ${\bf c}$, at each lattice site, ${\bf r}$. The time
evolution of  $n({\bf c},{\bf r},t)$ is described by the discretized
analog of the linearized  Boltzmann equation~\cite{succi}, involving
propagation and collision. Collisions are specified in order that the  hydrodynamic fields satisfy the Navier-Stokes equation at large scales. We will use units so that the  mass of the lattice molecule, the lattice spacing and the time step are unity, the kinematic viscosity $\nu$ of the fluid will be smaller than 1 and the speed of sound $c_s=1/\sqrt{2}$; the density $\rho$ is constant for practical purposes.  The suspended, spherical, particles of radius $R=5/2$   are modeled by the set of boundary links which join fluid nodes and nodes interior to the particles~\cite{laddjsp}. Stick boundary conditions are applied through appropriate  kinetic rules on the boundary  links which determine the force the fluid exerts on the particle; the equations of motion of the particles are then integrated using the self-consistent  method described in Ref.~\cite{lowe96}. Whenever two particles are at a distance slightly larger  than their diameter, an elastic collision is performed to ensure they do not overlap.

Since we are interested in the basic mechanisms controlling the collective dynamics of spps'  suspensions, we introduce a simplified self-propulsion mechanism, complementary to others proposed in the literature~\cite{sunil}. We subtract a constant amount  of momentum with  fixed magnitude $\Delta p_0$, uniformly  distributed over all the  fluid nodes connected to solid nodes and which lie  within a cone
of angle $\psi_0$ around a predefined  direction of motion. The value
of $\psi_0$ affects the details of the local flow field, but does not modify the collective features we will analyze;   for simplicity's sake we have chosen $\psi_0=\pi/2$. The direction of motion is determined at the startup and moves rigidly with  the propeller~\cite{llopis05}; in particular a spp rotation leads to the corresponding modification in its propulsion direction. The subtracted momentum is added to the spp, preserving momentum conservation. Since we are interested in the active motion of small organisms and colloids,   $\Delta p_0$ is  chosen to ensure that inertia is negligible~\cite{reynolds}.  A particle released from rest will reach  a  constant velocity inducing a dipolar flow in the embedding fluid at long distances.  Even if this mechanism is inspired by internal peristaltic motion~\cite{ajdaristone}, we will concentrate on generic aspects of propulsion. To this end, we will address the relevance of the symmetry of the fluid flow surrounding the  spps by comparing  this asymmetric driving, with a second symmetric propelling mechanism.  In order to focus on the basic mechanisms generated by the effective dissipative interactions induced by the fluid we disregard thermal fluctuations. 

\section{Short time dynamics}
A spp  of radius $R$ released from rest in  a fluid of viscosity $\nu$ will start moving and reach a terminal
constant velocity, $u_\infty$, on a characteristic  time $\tau_r\sim R^2/\nu$ in which  the fluid flow induced by the spp is established.  Consistent with this hydrodynamic origin, we have verified that the asymptotic velocity is reached algebraically, controlled by the long-time tail, $u(\tau)-u_\infty\sim\tau^{-d/2}$ ($d > 1$), $d$  being  the system's dimensionality. The asymptotic velocity arises from a balance between the driving force
and the friction, $u_{\infty}=\Delta p_0/(12 \pi \eta  R)$, where $\eta =\rho \nu$ is the solvent viscosity~\cite{llopis05}.

\begin{figure}
\includegraphics[scale=0.25,angle=-90]{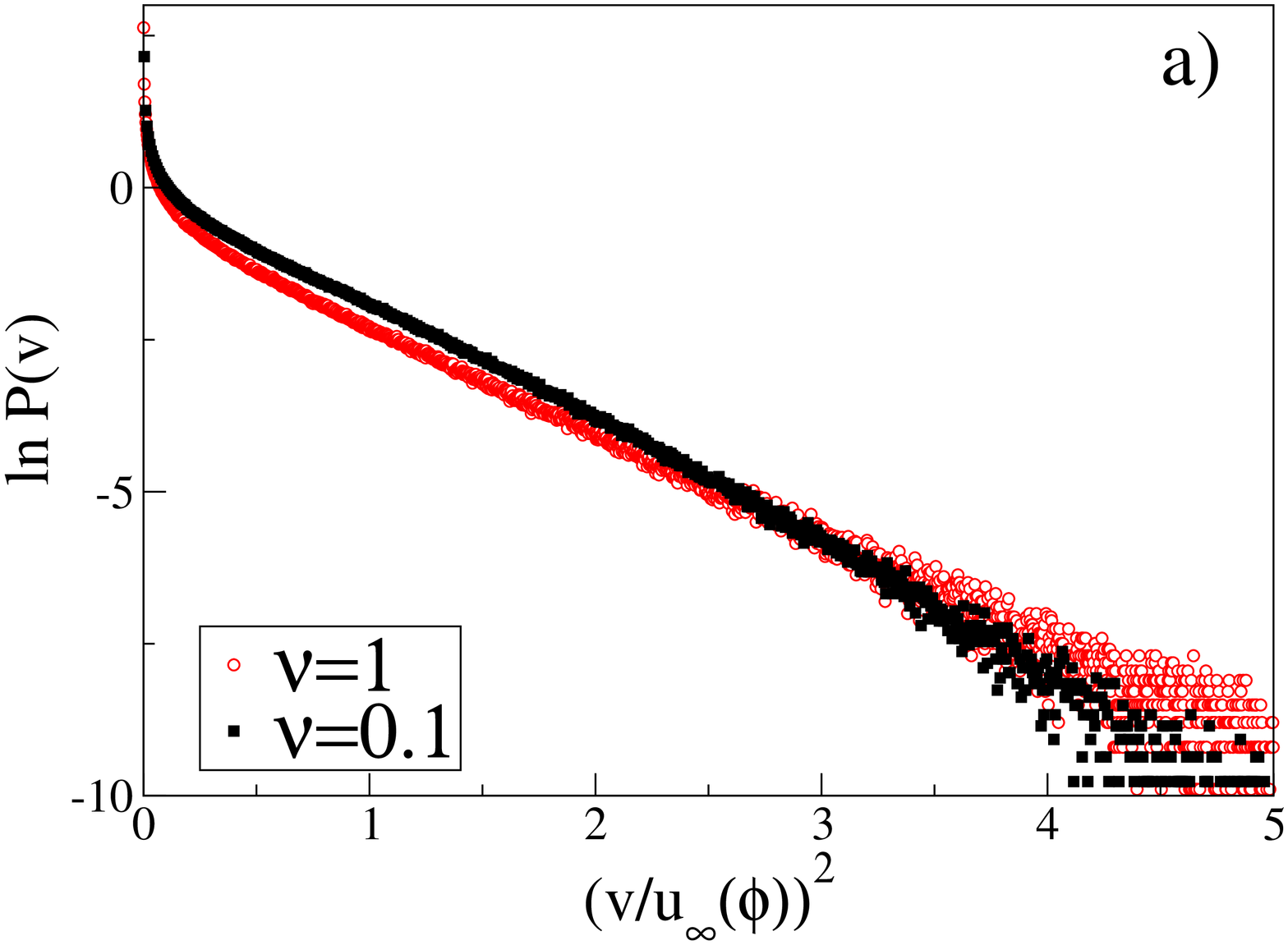}
\includegraphics[scale=0.25,angle=-90]{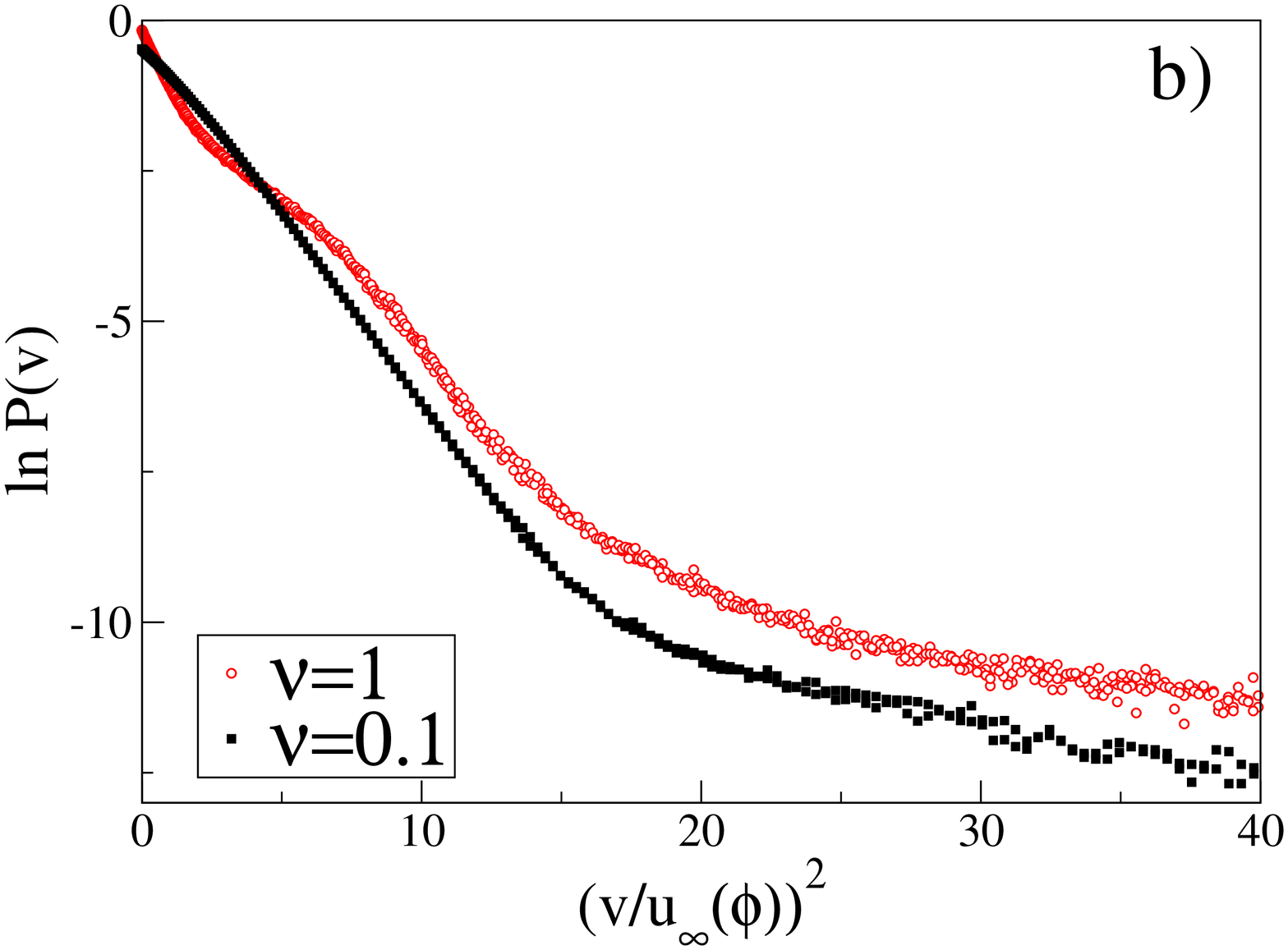}
\caption{Velocity distribution functions  at volume fraction
  $\phi=0.02$ in a system of $N=1020$ spps. Velocities are  collected from
  independent initial conditions at $\tau/\tau_r\sim 2$.  a) Particle
  velocity distribution function; b) fluid velocity distribution
  function.}
\label{fig:pdfshort}
\end{figure}

The induced flows generated by the propellers have a strong influence on their velocity  distributions; the non-equilibrium
character of their motion leads to velocity distributions that depart
markedly from a Gaussian, as displayed in
Fig.~\ref{fig:pdfshort}.a. The velocity distribution function, $P(v)$,
is characterized by an algebraic decay at small velocities, consistent
with $P(v)\sim 1/v$, while its asymptotic decay is consistent with a
Gaussian. In Fig.~\ref{fig:pdfshort}.b we show the corresponding fluid
velocity distribution; in this case the small velocity regime decays
exponentially,  followed by a decay compatible with a Gaussian in the
region of velocities of the order of  $u_\infty$ before it develops an
asymptotic decay characterized by a generalized exponential $P(v)\sim
\exp[-(v/v_0)^\alpha]$ with exponent $\alpha=0.4\pm0.2$ ~\cite{asymptotic}. Although deviations from Gaussian
behavior are known in other non-equilibrium systems, e.g. driven granular fluids~\cite{grains}, in spps suspensions they arise due to coupling to the collective motion of the embedding solvent.
\newline
 Interactions among  spps modify their effective mean asymptotic velocity at time scales  $\tau_r$, as depicted in
Fig.~\ref{fig:vphi},  where averages over 20 independent configurations have been carried out.  Spps  with parallel orientations interact more strongly than their randomly moving counterparts. The former situation is reminiscent of sedimentation in passive colloids and a  quantitative comparison, as displayed in Fig.~\ref{fig:vphi}  shows that the absence of a net applied external force leads to a weaker coupling, due to the dipolar nature of the  induced velocity flow~\cite{anderson}.
\section{Long time dynamics}
For the small $Re$ of interest,  spps' propulsion leads to configurational changes on time scales  $\tau_m \sim R/u_\infty > \tau_r$, when spps will also modify their  motion through  collisions.  We have observed that   angular velocity autocorrelation functions decay on  $ \tau_m$; hence  spps modify their direction of motion mostly through collisions. Although additional mechanisms which affect the spps' direction of motion, such as  tumbling,  play a relevant role in the decorrelation of particles' orientation, most of the features described in this manuscript survive  qualitatively if they are accounted  for~\cite{llopis05}. 
\begin{figure}
\includegraphics[scale=0.245,angle=-90]{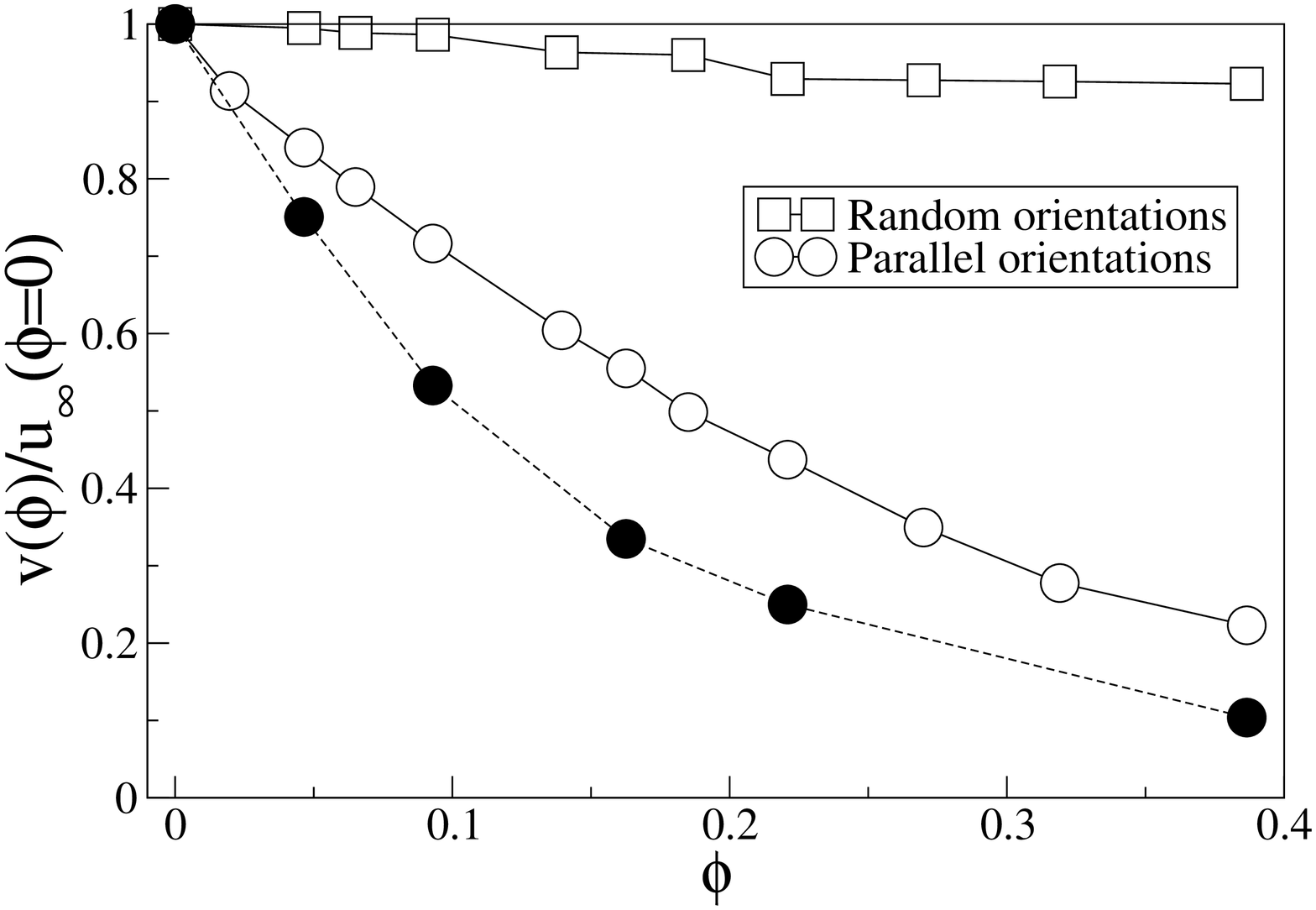}
\includegraphics[scale=0.245,angle=-90]{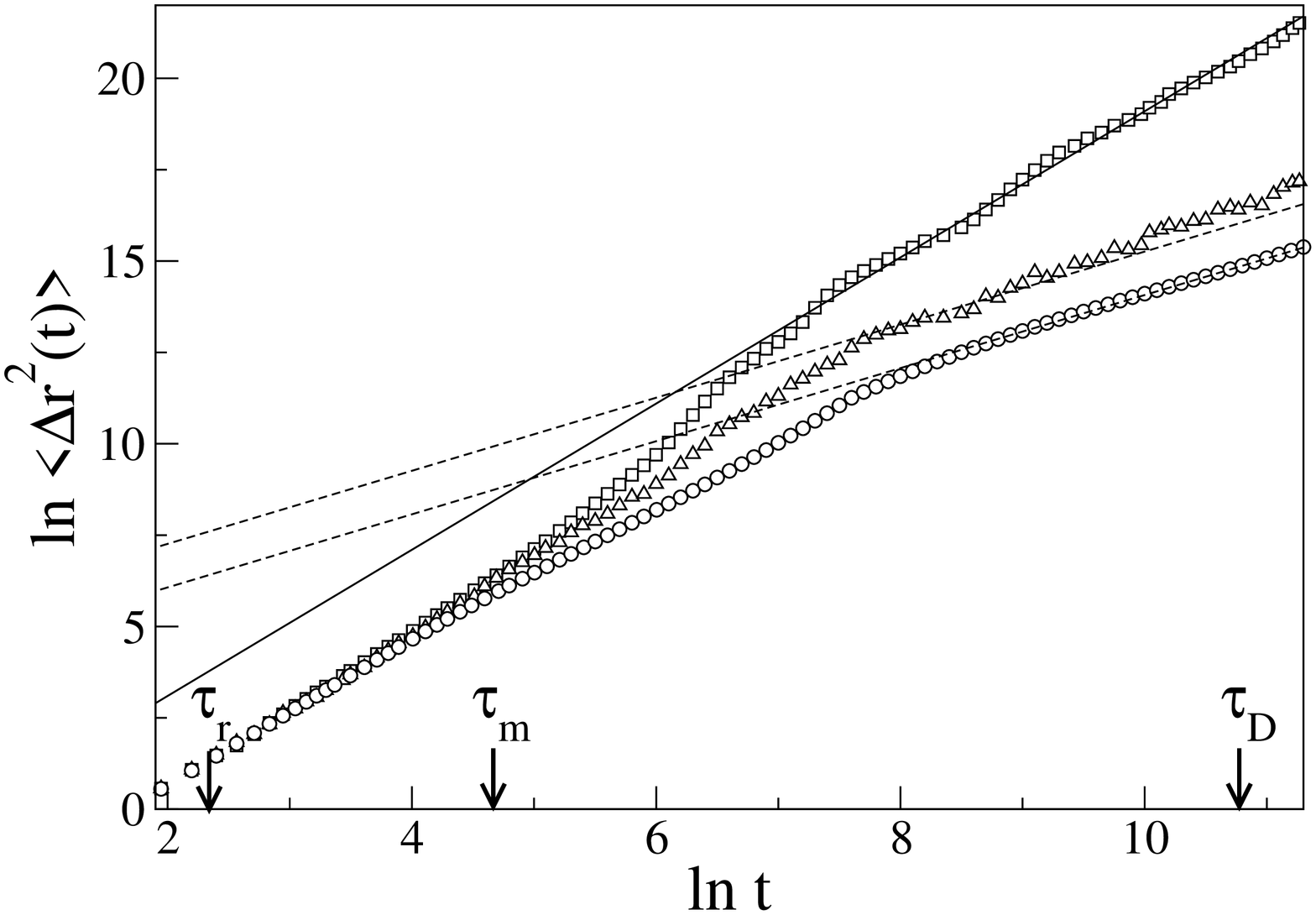}
  \caption{ Dependence of spps' velocities on volume fraction for
    random and parallel orientations for a system of $N=400$ spherical particles
    in a fluid of viscosity $\nu=0.5$.  Black circles: sedimentation velocity for hard spheres; it has been obtained with the same model by introducing a uniform external field and setting $\Delta p_0=0$.
  \label{fig:vphi}}
\caption{Mean square displacement in a spps' suspension for a
  two-dimensional system of $400$ propellers, kinematic viscosity
  $\nu=1$ and Reynolds $Re=0.25$. The continuous straight line
  corresponds to ballistic motion,  $\langle \Delta r^2 (t)
  \rangle\sim t^2$, while the  long-dashed ones correspond to
  diffusion, $\langle\Delta r^2 (t)\rangle \sim t$. Circles
  correspond to $\phi=0.282$ whereas  squares
  and triangles refer to $\phi=0.1$ for different initial conditions. We display also the characteristic
  times:  $\tau_r\sim R^2/\nu \sim 10$, $\tau_m \sim R/u_\infty \sim
  100$ and $\tau_D\geq R^2/D \sim 5\cdot10^4$.
    \label{fig:msd2D}}
\end{figure}
At  $\tau> \tau_m$ we enter in the long-time dynamics regime of spps suspensions. In order to  reach  this  long-time regime, we have performed computer simulations of the LB model in two dimensions~\cite{3d}. Fig.~\ref{fig:msd2D}  displays in circles the mean square displacement (msd)  at an intermediate volume fraction, $\phi=0.282$, with a crossover from ballistic to
 diffusive asymptotic behavior at the collision  time scale, $\tau\sim \tau_m$. At longer times, a  diffusive regime is achieved for  $\tau_D\geq R^2/D$, where $D$ is the diffusion coefficient. Since diffusion can only be achieved through succesive collisions and relaxations to the local fluid flows, $\tau_D > \tau_m$, which implies that the propellers' P\'eclet number satisfies $Pe\equiv
 u_\infty R/D > 1$; hence for these systems convection is always a relevant mechanism. This feature is consistent with the long crossover regime observed in the msd of beads in a bacteria suspension~\cite{libchaber00}.
 
At low volume fractions a different scenario is observed, as displayed
in squares and triangles in Fig.~\ref{fig:msd2D}. In the collision regime,
$\tau_m$,  a clear acceleration of the particles is observed; such
behavior leads at longer times to   a   superdiffusive regime. The observed acceleration at intermediate times is due to the increase in the spps' velocities through hydrodynamic coupling. We have seen that  collisions tend to correlate locally  spps  which move  initially at random. This tendency leads to an acceleration and to the formation of  large
clusters which eventually split into smaller
aggregates with a finite life time of the order of $\tau_m$. For some initial conditions, we have observed the appearance of system-spanning clusters that lead to long-time ballistic motion. The analysis of the propellers' velocity autocorrelation function (vacf) shows a consistent behavior; after  an initial exponential decay at times of order $\tau_m$, the vacf increases toward a plateau value at later times. The vacf shows small  oscillations in this late regime that we attribute to  cluster dynamics. These oscillations are also visible in the mean square displacement in Fig.~\ref{fig:msd2D}.a.
\begin{figure}
\includegraphics[scale=0.25,angle=-90]{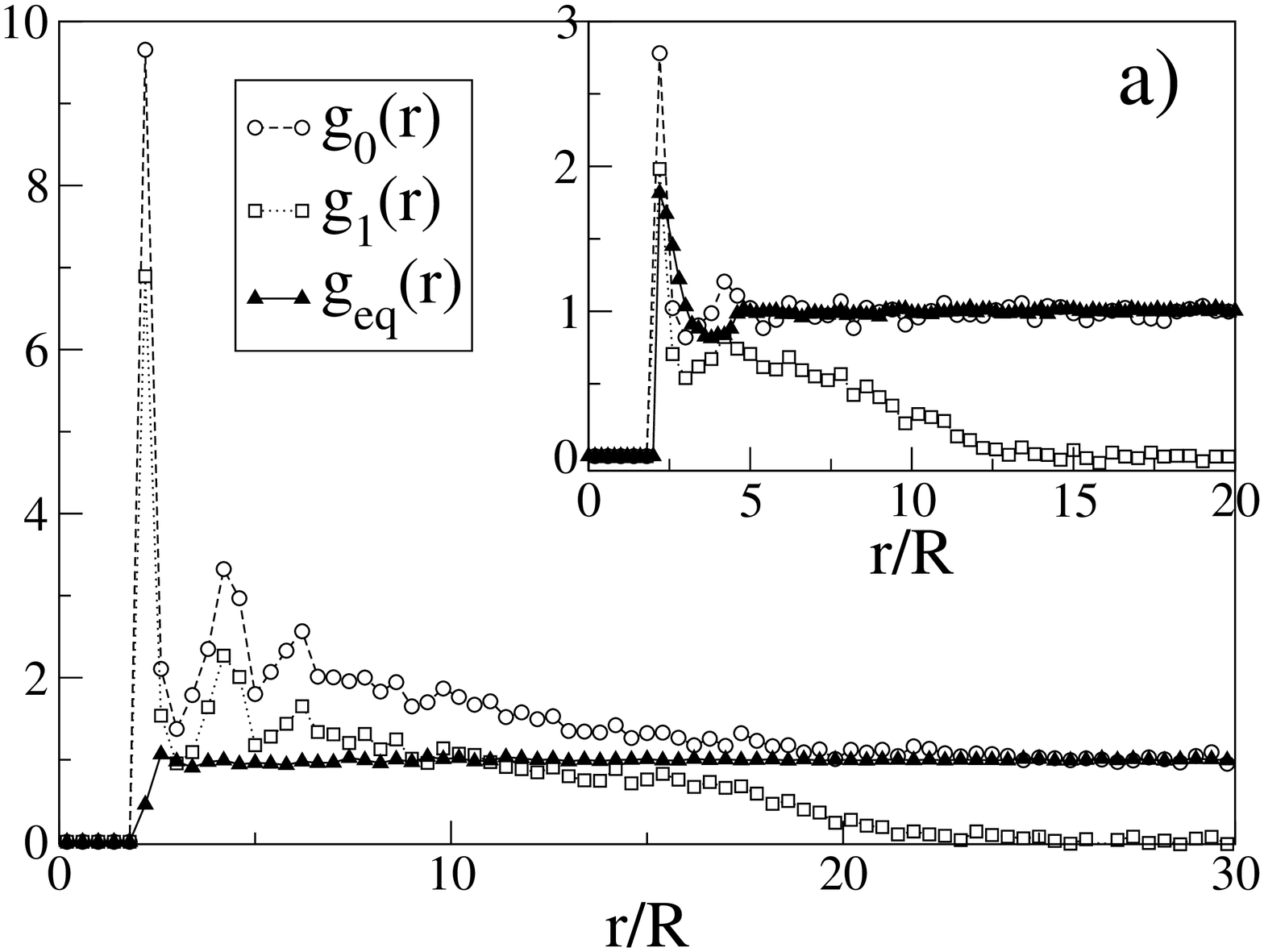}
\includegraphics[scale=0.25,angle=-90]{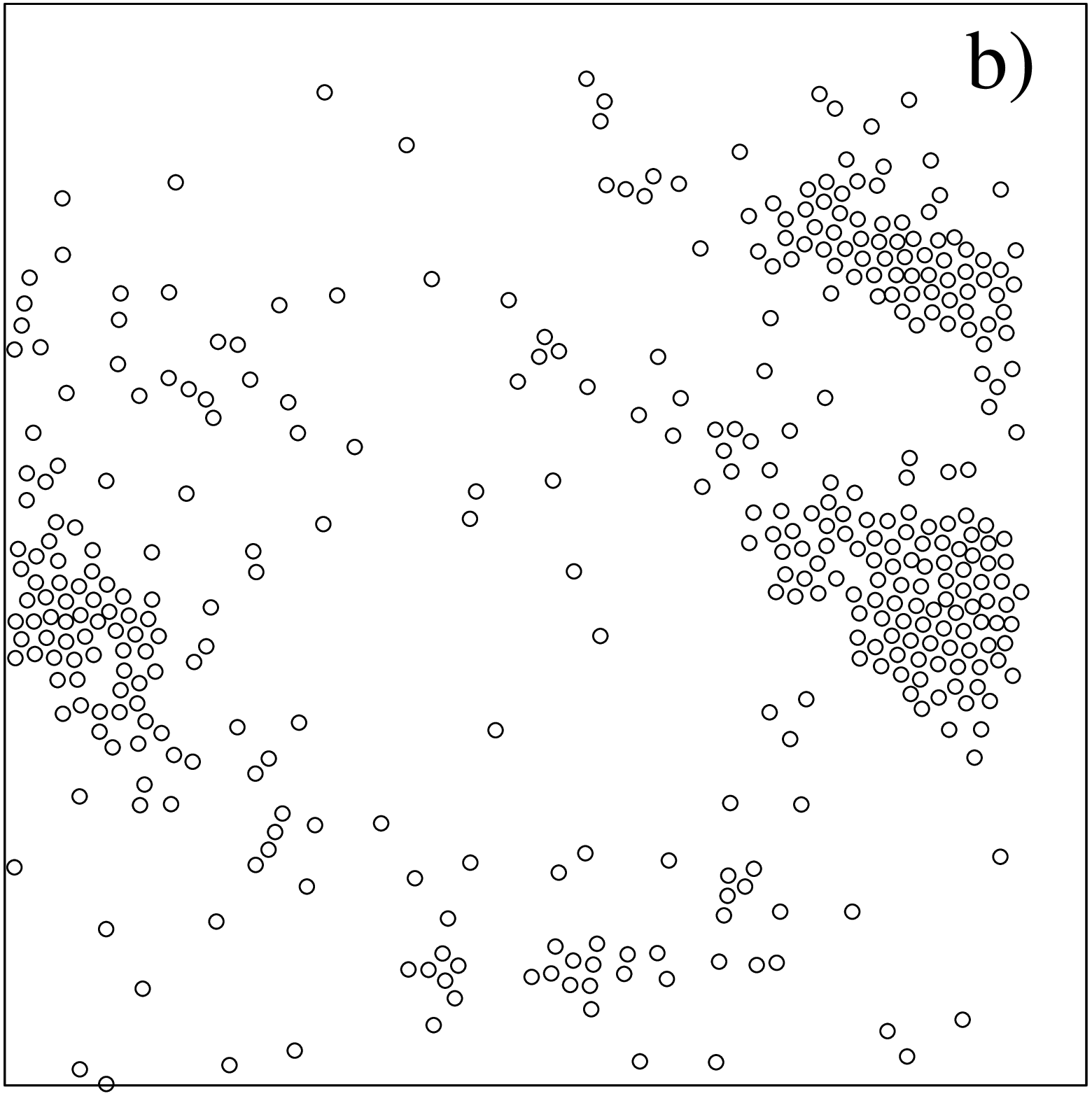}
\caption{a) Radial distribution functions of  a spps' suspension  for a two-dimensional system of $N=815$, kinematic viscosity $\nu=1$ and Reynolds $Re=0.25$ at times $\tau\geq 10 \tau_m$ and area fraction $\phi=0.1$; Inset: Radial distribution functions for an analogous system at   $\phi=0.1$, when the driving mechanism is symmetric around the direction of motion. See text for definition of $g_n$. Thick black line corresponds to the short time regime,  $\tau \sim \tau_r$.  b) Snapshot of the spps solution with asymmetric driving at $\phi=0.1$ at time $\tau\geq 90 \tau_m$. \label{fig:gr2D}}
\end{figure}

To gain insight in the structures the propellers develop, we have
analyzed their spatial distribution through  the  generalized  radial
distribution functions $g_n(r)\equiv\langle P_n(
\cos\theta_{ij})\rangle$, with $\theta_{ij}$ being the relative angle
between the   direction of motion of the  reference particle $i$ and
that of all particles $j$ at a distance between $r$ and  $r+dr$, where
$P_n$ is the n-th degree Legendre polynomial. Fig.~\ref{fig:gr2D}.a
shows the lowest order radial distribution functions,  where simulations over 5 independent initial configurations are used to increase  the statistical accuracy. The value of the
pair distribution function at contact, $g_0(r=2 R)$, is much greater
than its equilibrium  counterpart, which shows the tendency of the
spps to  remain closer as a result of their activity. One can see a
clear structuring on length scales much larger than the  spps' size
signalling  the  formation of  mesoscale structures.  The decay of  $g_0(r)$ is
consistent with the aggregates observed in simulations; in
Fig.~\ref{fig:gr2D}.b we show a typical configuration in the long-time
regime. The decay of  $g_1(r)$ on the same length scale, indicates
that there is a local tendency  of spps  to share common direction, so
that the transient aggregates have a common direction of motion. This implies
that spherical spps develop orientational order; their dynamics  will
display similarities with that of  non-spherical spps (see
e.g.~\cite{ramaswamy04}).  The observed  aggregates  form and
dissolve, and hence there is no sign of  permanent aggregation induced
through hydrodynamics. Additional interactions, for example chemical signaling, may  stabilize these aggregates. We have also analyzed the velocity distribution functions both for propellers  and fluid  at long times, $\tau\geq \tau_D$, where we have not found significant deviations from Gaussian behavior.

We have also studied the collective behavior of
symmetric  propellers to assess the relevance of the  specificity of the propulsion
mechanism.  In this case the momentum, $\Delta p_0$, is
extracted uniformly from all nodes lying within the two cones defined
by the angles $\psi_0$ and its suplementary, $\pi-\psi_0$. For this
driving the velocity distributions at short times are equivalent to
those found for asymmetrically driven  propellers. Asymptotically
($\tau\geq \tau_D$), we have  observed  diffusive behavior for all
the concentrations explored. Correspondingly,  the radial distribution function $g_0(r)$ does not show a slow decay
on distances larger than $R$, as displayed in the inset of
Fig.~\ref{fig:gr2D}.. In this case one do not observe  transient, mesoscopic, aggregates.  The contact value is still significantly larger than its equilibrium counterpart, although not as large as for asymmetric driving. In this case  $g_1(r)$ decays to zero on longer length scales than $g_0(r)$, which shows a tendency to local orientation of spps, and hence a sensitivity toward nematization, as has been predicted for non-s`herical spps~\cite{ramaswamy04}. Therefore,  hydrodynamics  makes particles approach, while its coupling strength determines the   properties of spps' structures. 

Even if we cannot rule out a further relaxation to an asymptotic diffusive regime for asymmetric propulsion at low $\phi$, the observed  formation and dissolution of aggregates on the characteristic collision time scale, $\tau_m$, and the corresponding oscillations in the vacf strongly suggest that the superdiffusive regime is properly sampled. We have checked that this  long-time  behavior  is robust upon  changing the system size and for a variety of volume fractions. We have also seen that the dependence of the msd  on volume fraction is consistent; i.e. upon increasing $\phi$ we observe superdiffusive behavior, then a crossover region where the dynamics is intermediate on the simulation time scale which ends up becoming diffusive  at higher volume fractions.

The transition to collective motion, characterized by the asymptotic superdiffusive regime, is analogous to reported flocking transitions~\cite{toner}, although  the  mechanism  in this case is dissipative, through hydrodynamics. We have verified that this transition disappears when considering a symmetric  propulsion mechanism, and hence  attribute  flocking as arising from the strength of hydrodynamic coupling due to the  drag a moving spp induces on neighboring propellers. Such a dissipative interaction is  responsible both  for the velocity increase and for spps' alignment observed at $\tau \geq \tau_m$. In previous flocking models, the transition to collective  macroscopic motion is determined by the effect of interactions on  the direction of motion. These interactions  are  absent in our model, which indicates that hydrodynamic coupling will favor previously observed flocking transitions once the additional interactions between propellers are accounted for, and will promote it at smaller volume fractions.  
  
\section{Conclusions}
We have introduced a new, simple mesoscopic model for self-propulsion which resolves  hydrodynamics and  have identified the relevant dynamical regimes for low Reynolds numbers spps suspensions.  In this
model we have been able to carry out a detailed analysis which has allowed us to clarify  the role of hydrodynamics on the different  time scales. 

In particular, we have considered the correlations induced by these dissipative interactions at short times, which lead to  clear deviations from  Gaussian velocity distributions. Such effects are visible both in the spps and in the  fluid velocity distributions. Spps interact  through collisions at longer time scales and hydrodynamics can accelerate them and favor the formation of transient aggregates. Depending on the details of the propulsion mechanism, spps may exhibit both diffusive and superdiffusive dynamics at the scales at which structural rearrangements take place. The  understanding of these regimes, their connections to flocking transitions and their implications in the collective properties of spps suspensions require a more  detailed study.

\acknowledgments
The authors acknowledge financial support from DGICYT of the Spanish Government, they  thank R. Adhikari and M.E. Cates for fruitful discussions, and M.E. Cates for suggestions and a careful reading of the manuscript. I.Ll. wants to thank the HPC-EUROPA project (RII3-CT-2003-506079), with the support of the European Community - Research Infrastructure Action. I.P. thanks Distinci\'o from DURSI (Generalitat de Catalunya) and P.B. Sunil Kumar for fruitful discussions.

\end{document}